\documentclass[11pt]{article}
\usepackage{moriond-ms,epsfig}
\usepackage{amssymb,latexsym}
\usepackage{amsmath}

\def\Journal#1#2#3#4{{#1} {\bf #2}, #3 (#4)}

\def\PRD{{\em Phys. Rev.} D}

\def\be{\begin{equation}}
\def\ee{\end{equation}}
\def\bea{\begin{eqnarray}}
\def\eea{\end{eqnarray}}

\def\lsim{\hbox{ \raise.35ex\rlap{$<$}\lower.6ex\hbox{$\sim$}\ }}
\def\gsim{\hbox{ \raise.35ex\rlap{$>$}\lower.6ex\hbox{$\sim$}\ }}
\def\lsim{\hbox{ \raise.35ex\rlap{$<$}\lower.6ex\hbox{$\sim$}\ }}
\def\gsim{\hbox{ \raise.35ex\rlap{$>$}\lower.6ex\hbox{$\sim$}\ }}
\def\setC{\mathbb{C}}
\def\setR{\mathbb{R}}

\begin{document}
\vspace*{4cm}
\title{CONSTRAINING SUSY GUTs WITH COSMOLOGY}

\author{MAIRI SAKELLARIADOU$^1$ and JONATHAN ROCHER$^2$}

\address{$^1$Division of Astrophysics, Astronomy, and Mechanics, Department
of Physics,  University
of Athens, Panepistimiopolis, GR-15784 Zografos, Hellas, and 
\\
Institut d'Astrophysique
de Paris, 98bis boulevard Arago Arago, 75014 Paris, France.\\
$^2$Institut d'Astrophysique
de Paris, {${\cal G}\setR\varepsilon\setC{\cal O}$} , FRE 2435-CNRS, 98bis
bld Arago, 75014 Paris, France.}

\maketitle\abstracts{Cosmic strings of the GUT scale, generically
formed during the SSB of supersymmetric hybrid inflation, are
compatible with the most recent CMB data. The strong constraints on the
allowed cosmic strings contribution to the measured temperature
anisotropies impose limits on the free parameters of the inflationary
models, namely the mass scales and the couplings.}

\section{Introduction}
According to the Grand Unified Theories (GUTs), topological defects are
expected to be formed during the phase transitions accompanied by
Spontaneously Symmetry Breaking (SSB), as the Universe cools down
during its expansion. Among the various kinds of local (gauge)
topological defects only cosmic strings are allowed; monopoles and
domain walls would overlcose the Universe, implying that an inflationary era
is required after the formation of the harmful defects. Considering
all SSB schemes from a large gauge group down to the standard
model, cosmic strings, occassionally accompanied by embedded strings,
are always left behind after the last inflationary era~\cite{rjs}.
However, constraints are also imposed on cosmic strings. More
precisely, cosmic strings of the GUT scale should not contribute more
than about $10\%$ to the CMB temperature anisotropies~\cite{bprs,p}.
This CMB constraint on cosmic strings does not imply that GUT scale
cosmic strings are ruled out. It can instead be used to constraint the
free parameters (mass scales and couplings) of the inflationary model. 
We study~\cite{jm1,jm2} these constraints for F-term and D-term 
inflation.

\section{F-term inflation}
F-term inflation is based on the SUSY renormalisable superpotential
\begin{equation}\label{superpot}
W_{\rm infl}^{\rm F}=\kappa S(\Phi_+\Phi_- - M^2)~,
\end{equation}
with $S, \Phi_+, \Phi_-$  three chiral superfields and $\kappa$,
$M$   constants.  Including the one-loop radiative
corrections to the scalar potential  along the inflationary valley,
the effective potential reads
\begin{equation}
\label{VexactF}
V_{\rm eff}^{\rm F}(|S|)=\kappa^2M^4\left\{1+\frac{\kappa^2
\cal{N}}{32\pi^2}\left[2\ln\frac{|S|^2\kappa^2}{\Lambda^2}+
(z+1)^2\ln(1+z^{-1})+(z-1)^2\ln(1-z^{-1})\right]\right\}~,
\end{equation}
where $z=|S|^2/M^2\equiv x^2$, and $\cal{N}$ stands for the dimensionality
of the representations to which the complex scalar components $\phi_+,
\phi_-$ of the chiral superfields $\Phi_+, \Phi_-$ belong. Now, $S$ denotes
 the scalar component of its superfield.

\begin{figure}
\rule{5cm}{0.2mm}\hfill\rule{5cm}{0.2mm}
\psfig{figure=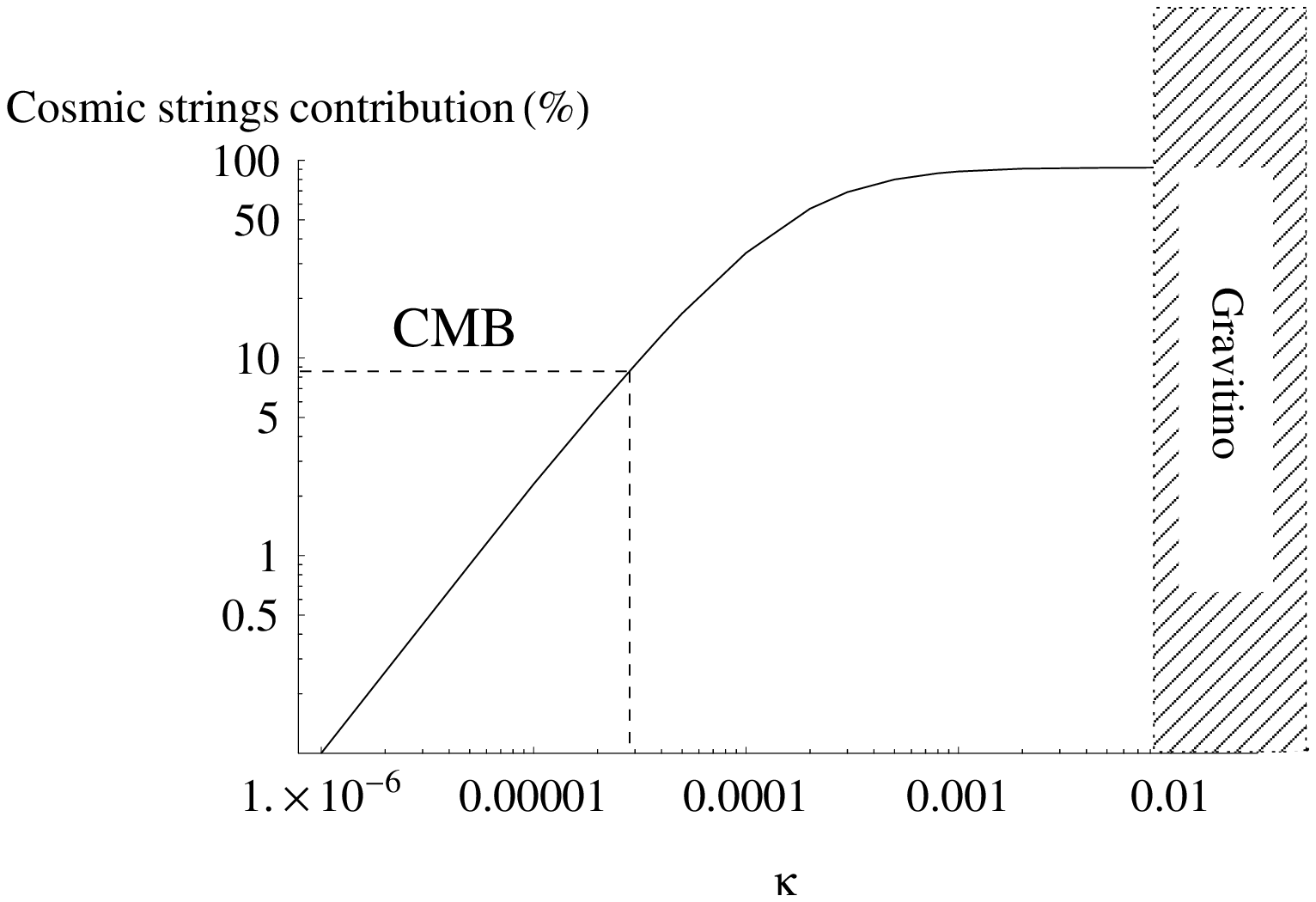,height=2.in}
\psfig{figure=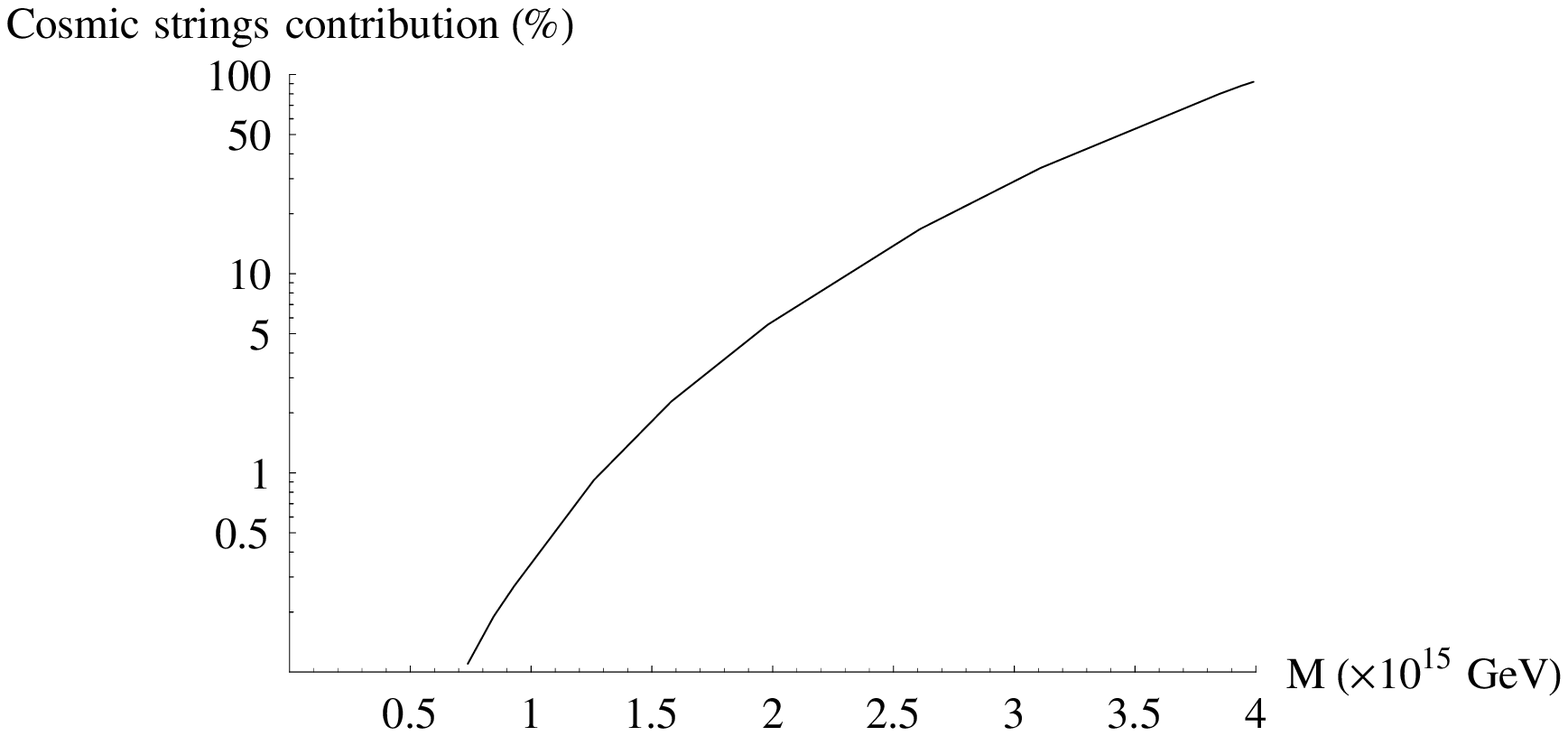,height=2.in}
\caption{On the left, constraints on the single parameter $\kappa$ of the
  model (for $\mathcal{N}=\mathbf{3}$).  On the right, evolution of the
  cosmic strings contribution to the quadrupole anisotropy as a
  function of $M$. 
}  \rule{5cm}{0.2mm}\hfill\rule{5cm}{0.2mm}
\label{figa}
\end{figure}

On large angular scales, the main contribution to the CMB anisotropies
is given by the Sachs-Wolfe effect. The quadrupole anisotropy,
normalized to the DMR-COBE data, has two contributions: one coming
from the quantum fluctuations of the inflaton field and another one
from the cosmic strings network. The two contributions can be computed
numerically for given values of $\kappa, \mathcal{N}$ and fixing the number
of e-foldings.  We find~\cite{jm1,jm2} that the mass scale $M$ grows
very slowly with $\cal N$ and it is of the order of $10^{15}$ GeV. The
parameter $M$ sets the scale of inflation, and the mass scale
of cosmic strings formed at the end of the inflationary era. The CS
contribution to the CMB is a function of the coupling $\kappa$, or
equivalently of the mass scale $M$. The CS contribution to the WMAP
measurements should not be higher than $9\%$, which implies
\begin{equation}
\kappa \lsim 3\times10^{-5} \times \frac{3}{\mathcal{N}} {\rm ~with~}
\mathcal{N} \in 
\{\mathbf{1}, \mathbf{3},\mathbf{16},\mathbf{126}\}
~~\Leftrightarrow ~~M \lsim 2.2\times 10^{15}~{\rm GeV}~ {\rm for~ ~all~}
\mathcal{N}~.
\end{equation}
The gravitino constraint on the reheating temperature results to a
weaker upper bound for $\kappa$, namely $\kappa \lesssim 8.2\times10^{-3}$.
The CMB and gravitino constraints on $\kappa$, as well as the cosmic
strings contribution to the CMB data as a function of $\kappa$, are
given in Fig.~\ref{figa}.

\begin{figure}
\rule{5cm}{0.2mm}\hfill\rule{5cm}{0.2mm}
\psfig{figure=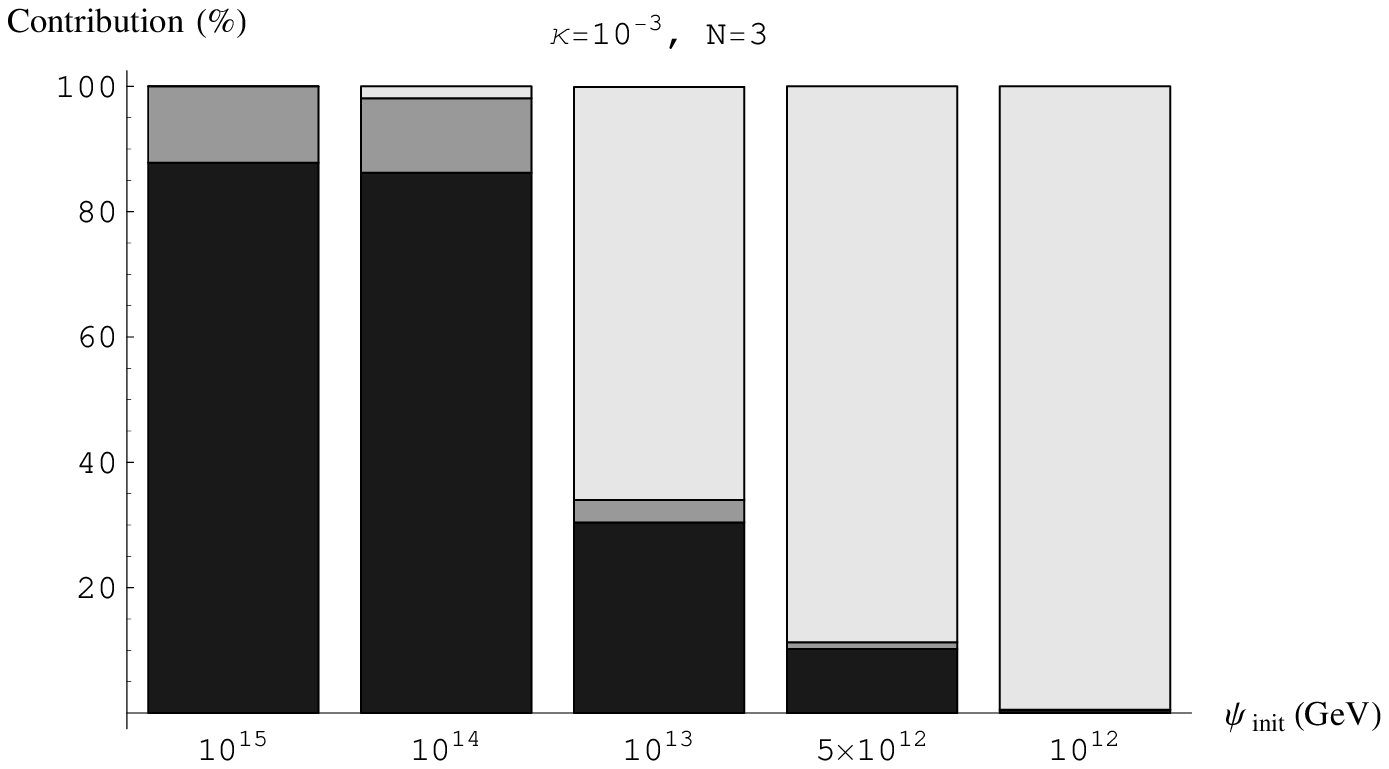,height=2.in}
\psfig{figure=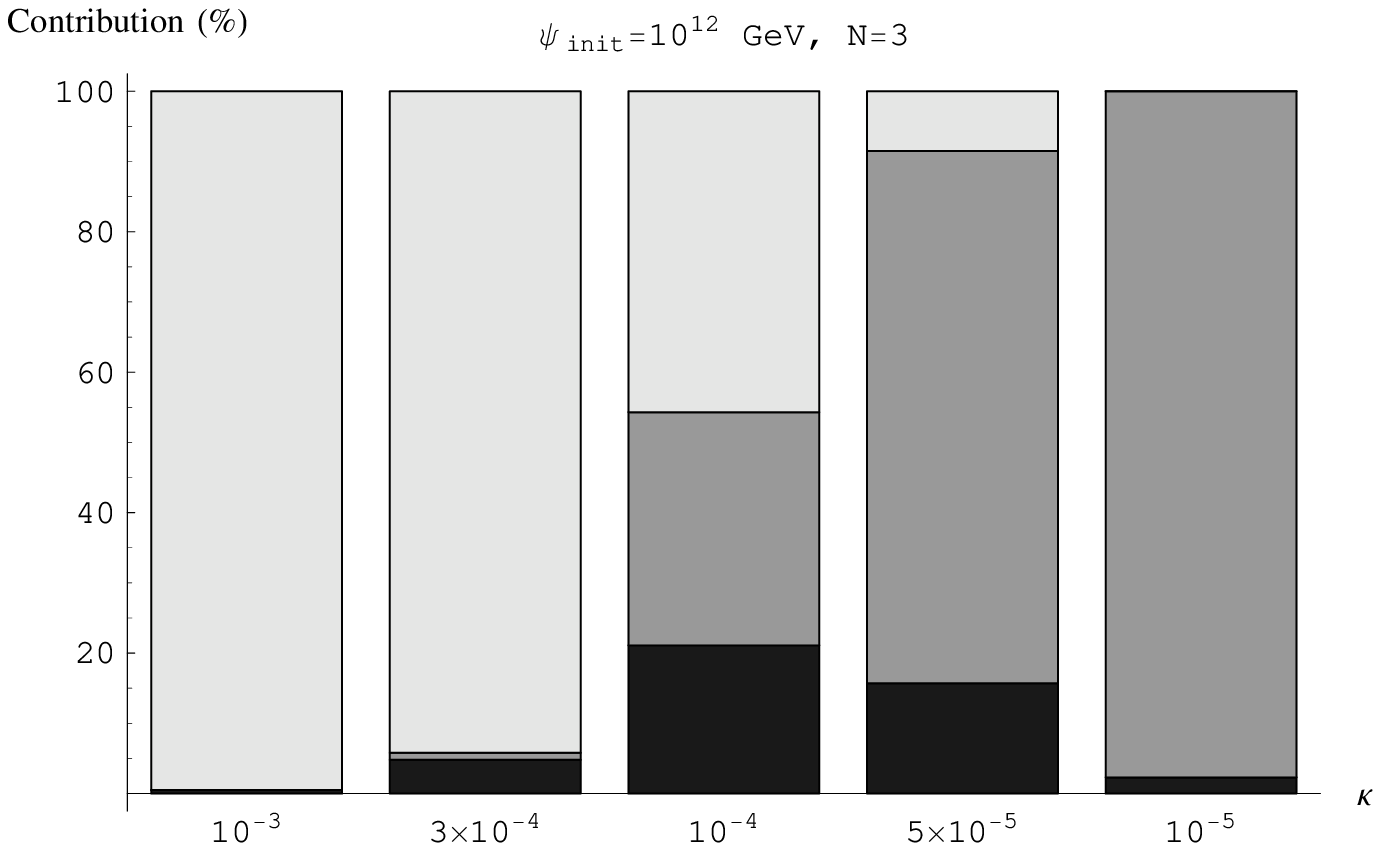,height=2.in}
\caption{The cosmic strings (dark gray), curvaton (light gray) and
inflaton (gray) contributions to the CMB  as a
function of the the initial value of the curvaton field
${\cal\psi}_{\rm init}$, and the superpotential coupling 
$\kappa$, for ${\cal N}=3$.}
\rule{5cm}{0.2mm}\hfill\rule{5cm}{0.2mm}
\label{contribcurv}
\end{figure}

The allowed upper limit on the coupling $\kappa$ can be increased if we 
employ the curvaton mechanism, in which case we have one extra parameter, 
namely the initial amplitude of the curvaton field, $\psi_{\rm init}$.
The CMB data can be used to constrain the upper bound on $\psi_{\rm init}$,
which depends on the coupling $\kappa$. We find $\psi_{\rm init}\lsim
5\times 10^{13}(\kappa/10^{-2}) {\rm GeV}$, for $\kappa$
in the range $[5\times 10^{-5},1]$;
smaller values lead to a CS contribution  below the WMAP limit.
These
findings are summarized in Fig.~\ref{contribcurv}.
The Supergravity (SUGRA) corrections can be neglected in the framework
of F-term inflation and this is expected since the inflaton field is
below the Planck mass by three orders of magnitude.

\section{D-term inflation}

D-term inflation is derived from the superpotential
\begin{equation}
\label{superpotD}
W^{\rm D}_{\rm infl}=\lambda S \Phi_+\Phi_-~.
\end{equation}
D-term inflation requires the existence of a nonzero Fayet-Illiopoulos
term $\xi$, permitted only if an extra U(1) symmetry is introduced.
D-term inflation must be studied in the framework of SUGRA~\cite{jm1,jm2}, 
since an
analysis within local SUSY results to an inflaton field of the order
of the Planck mass or higher.

For minimal supergravity, the effective potential keeping all trems of
the one-loop radiative corrections read~\cite{jm1,jm2}
\begin{equation}
\label{VexactDsugra}
V^{\rm D}_{\rm eff} =
\frac{g^2\xi^2}{2}\left\{1+\frac{g^2}{16\pi^2}
\left[2\ln\frac{|S|^2\lambda^2}{\Lambda^2}\exp\left({|S|^2\over M_{\rm
Pl}^2}\right)+
(z+1)^2\ln(1+z^{-1})+(z-1)^2\ln(1-z^{-1})\right]\right\}~,
\end{equation}
where $z=[\lambda^2 |S|^2/(g^2\xi)]\exp(|S|^2/M_{\rm Pl}^2)\equiv
x^2$.  The constant term in the potential is identical to the SUSY
case, but its first derivative is modified by the factor
$(1+|S|^2/M_{\rm Pl}^2)$. D-term inflation is trickier
than F-term, but it is still possible to solve numerically and find the
various contributions to the CMB data. 

We note that in the above effective potential we do not consider
quantum gravitational corrections, since their computation results to
a negligible contribution to the effective potential.

We find that the inflaton field is of the order of $10 M_{\rm Pl}$,
for the studied parameter space in $\lambda$ and $g$.  The cosmic
strings contribution to the quadrupole anisotropy is not constant, nor
is always dominant, in contradiction to previously made statements. It
depends strongly on the value of the gauge coupling $g$ and the
superpotential coupling $\lambda$.  We find that $g\gtrsim 1$
necessitates multiple-stage inflation, since otherwise this inflationary era cannot last for 60 e-foldings, while $g\gtrsim 2\times 10^{-2}$ is incompatible with
the allowed CS contribution to the CMB measurements.  For $g\lesssim
2\times 10^{-2}$, the CMB constaint sets an upper bound to the allowed
window for $\lambda$, namely $\lambda \lesssim 3\times 10^{-5}$. 
SUGRA corrections result to also a lower bound on $\lambda$, which however 
is very small.
 
\begin{figure}
\rule{5cm}{0.2mm}\hfill\rule{5cm}{0.2mm}
\psfig{figure=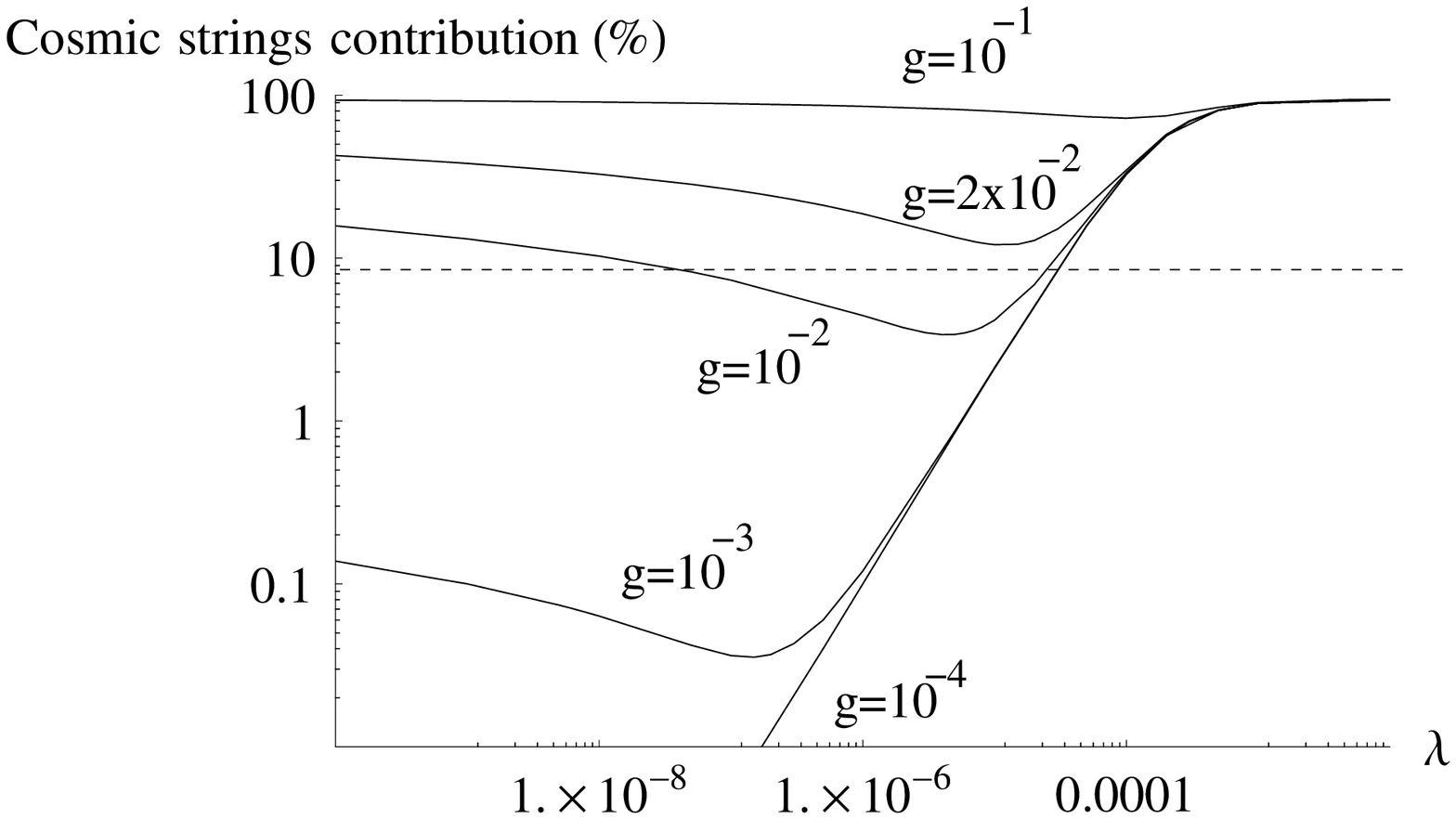,height=2.in}
\psfig{figure=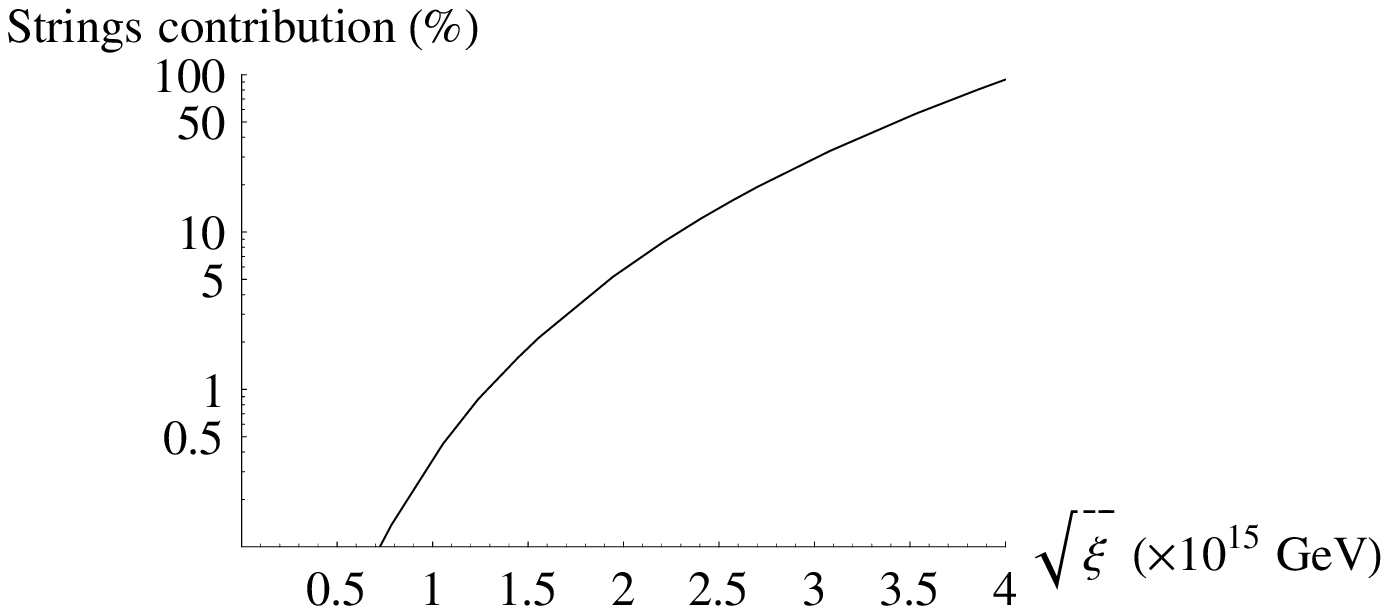,height=2.in}
\caption{On the left, 
cosmic strings contribution to the CMB temperature
anisotropies, as a function of the superpotential coupling  $\lambda$,
for different values of the gauge coupling $g$. The maximal contribution 
alowed by WMAP is represented by a dotted line. On the right, cosmic strings contribution as a function of the
mass scale $\sqrt\xi$. This holds for all studied values of $g$.}
\rule{5cm}{0.2mm}\hfill\rule{5cm}{0.2mm}
\label{PRL4}
\end{figure}

The CMB constraint on the couplings of the superpotential can be also expressed
as a constraint on the mass scale, which in D-term inflation is given by the 
Fayet-Iliopoulos term. This constraint reads
\begin{equation}
\sqrt\xi\lesssim 2.3\times 10^{15}{\rm GeV}~.
\end{equation}
Our results are summarised in Fig.~\ref{PRL4}.

\section{Conclusions}
CMB measurements allow a small but non negligible contribution of
cosmic strings to the temperature anisotropies. This results to
constraints on the free parameters of supersymmetric hybrid inflation.
We thus put bounds on the mass scales and couplings of the superpotential.
Our study shows that both F-term as well as D-term inflationary models
are open possibilities. Thus, we clearly disagree with previous
statements that D-term inflation is ruled out.

\section*{Acknowledgments}
It is a pleasure to thank the organizers of the Moriond meeting for
inviting me to give this talk.

\end{document}